# Effect of alkyl substituents on the adsorption of thienylenevinylene oligomers on the Si (100) surface

B. Grandidier[a,]*, J.P. Nys[a], D. Stiévenard[a], C. Krzeminski[a], C. Delerue[a], P. Frère[b], P. Blanchard[b], J. Roncali[b]

[a] *Institut d'Electronique et de Microélectronique du Nord, IEMN, (CNRS, UMR 8520) Département ISEN, 41 bd Vauban, 59046 Lille Cédex (France)*

[b] *Ingénierie Moléculaire et Matériaux Organiques, CNRS UMR 6501, Université d'Angers, 2 bd Lavoisier, 49045 Angers (France)*

**ABSTRACT**

The adsorption of thienylenevinylene oligomers on the Si(100) surface has been investigated using scanning tunneling microscopy. The mode of substitution of the thiophene ring exerts a strong influence on the adsorption configurations and the images of the oligomer based on 3,4-dihexyl thiophene are highly voltage dependent. We discuss the influence of the alkyl chains on the adsorption process and on the appearance of the molecules in the STM images.

* Corresponding author: phone: +33 3 20 30 40 54; fax: +33 3 20 30 40 51

e-mail: grandidier@isen.iemn.univ-lille1.fr

# I. INTRODUCTION

π-Conjugated oligomers are subject to intense research activity due among others to their potential use as molecular wires in future molecular electronic devices [1]. For example, the characterization of the electron transfer through short chain oligothiophenes connected to metallic electrodes has been reported [2]. As microelectronics technology is based on the use of silicon substrates, there is an increasing need to link organic molecules to the existing silicon technology.

The optimization of the electronic transfer between the molecules and the silicon surface implies a detailed understanding of the nature of the interface between the molecules and the silicon surface.

In recent years, adsorption of unsaturated organic molecules on the Si(100) surface in ultra-high vacuum (UHV) has been studied and has revealed the possibility to attach molecules to the surface in a controlled manner [3]. Indeed, the silicon (100) surface is made up of silicon dimer rows. The bonding between the two adjacent Si atoms of a dimer can be described in terms of a strong σ and a weak π bonds. Due to the ease of the π bond cleavage, unsaturated organic molecules can chemisorb to the silicon surface by reactions similar to the reactions of cycloaddition, involving purely organic molecules in organic chemistry. These reactions lead to the creation of strong covalent bonds between the molecules and the silicon surface, resulting in the formation of a well-defined interface.

As shown in recent work, thienylenevinylene oligomers (nTVs) form a new class of π-conjugated oligomers of particular interest as molecular wires since they exhibit the largest effective conjugation and hence smallest HOMO-LUMO gap among extended oligomers with chain length in the 10 nm regime [4]. Since nTVs contain unsaturated double bonds, cycloadditions reactions with the Si(100) surface may be expected and would allow a good connection of these organic chains with the silicon surface. However, formation of bonds can occur on the entire length of the oligomers and thus could affect the π-electronic delocalization in the structure.

To gain a better understanding of the nTVs interaction with the Si(100) surface, we have studied the adsorption of 4TV oligomers. Since reactions of the silicon surface with this molecules are expected, two different 4TVs have been synthesised: the unsubtituted 4TV (Fig. 1a) and the 4TVH, where the positions 3 and 4 of the thiophene rings are substituted with hexyl chains (Fig. 1b). While both oligomers have the same electronic structure [4,5], this substitution inhibits the dimerization process [6]. Thus the substitution should only have an effect on the interactions of the molecules with their physical and chemical environment. We have used the scanning tunneling microscope to observe the different arrangements of the 4TV and 4TVH molecules after their adsorption on the Si(100) (2x1) surface. We show that the hexyl chains play a key role in the adsorption mechanism of the nTV molecules: 4TVs react strongly with the surface whereas 4TVHs interact weakly keeping the conjugation of the oligomer almost intact.

## II. EXPERIMENTAL

To check if the evaporation process does not alter the oligomers, they are first characterized by UV-vis spectroscopy in solution, then in the solid state, where they are adsorbed on glass. They are finally dissolved to measure again their UV-vis absorption in solution. Figure 2 shows the electronic absorption spectra obtained for the 4TV oligomers in a $CH_2Cl_2$ solution before and after the evaporation. Both spectra show the same peaks and this result indicates clearly that 4TVs can be evaporated without a modification of their structure.

Observations of the oligomers are performed with a scanning tunneling microscope (STM) in an UHV system. The n-type (5 Ωcm) Si(100) wafers are resistively heated to 600 °C for 12 h to degas the sample, cleaned by heating at 900 °C and flashing at 1280 °C for several seconds. Prior to the adsorption of the molecules on the Si surface, the surface is observed in scanning tunneling microscopy (STM) to check its cleanness and the defect densities, which has to be small in comparison with the dose of molecules adsorbed on the surface. The 4TVs oligomers are deposited into a Mo crucible, which is transferred into an homemade evaporator in UHV. Before the deposition process, the oligomer powder is outgassed by heating for several hours. The evaporation process is controlled by a quartz crystal microbalance. During deposition, the distance between the evaporation source and the substrate was 3 cm. The source temperature is measured by a (Tungstene-Rhenium) thermocouple in close contact with the crucible. The evaporation temperatures of 4TVH and 4TV range from 150 °C to 180 °C. The substrate temperature is kept at room temperature during the deposition. The base pressure in the evaporation chamber is around $2 \times 10^{-9}$ Torr during the deposition and the deposition process lasted only a few seconds to get submonolayer coverages. At the end of the deposition, the sample was immediately transferred to the STM chamber, where the base pressure is below $5 \times 10^{-11}$ Torr. After the transfer of the sample, a mass spectrometer was used in the evaporation chamber to check the absence of decomposition products such as thiophene moieties or alkanes.

## III. RESULTS AND DISCUSSION

Figure 3 shows the Si(100) (2x1) surface after the adsorption of 4TV. The silicon dimers are visible, forming rows of grey bean shaped. On top of these rows, two different types of bright features can be seen. Feature A can be described as a rounded protusion, whereas feature B has an oval shape with the direction of elongation parallel to the Si=Si dimer bonds.

To characterize more accurately the adsorption sites of both features, we show, in Fig. 4, a high resolution STM image of the Si(100) surface obtained after the adsorption of the molecules. In this image, the Si atoms of the dimers are resolved separately. Their precise location allows the determination of the A and B feature adsorption sites. The type A feature appears to be centered over a single Si atom of a dimer. The B type feature is localized over a single dimer unit. Although adsorbates associated with the B features can be seen isolated, they tend to form arrays extending in a direction parallel to the dimer bonds, as shown more clearly in Fig. 3.

From the dimension of the 4TV given in Fig. 1, it is clear that the features A and B do not have the expected size of 2.5 nm in the (100) plane. Furthermore, the heights of features A and B are respectively 1.7 Å and 1.3 Å and thus a vertical positioning of the oligomers is ruled out. Alkenes react with a high probability with the Si(100) surface by reactions of additions. As the 4TV oligomers contain C=C bonds in the thiophene rings and between the thiophene rings, they are likely to react with the Si=Si dimers [7,8]. Figures 3 and 4 clearly show that the oligomers have lost their structure by reacting with the surface. Since the features A and B can be found isolated, it is highly probable that the oligomers are broken. Therefore the interaction between 4TVs and the Si(100) surface leads to a dissociation of the molecule.

While the adsorption of 4TV oligomers gives features with a small size, the adsorption of 4TVH oligomers is quite different. A low coverage STM image of the Si(100) surface after the deposition of 4TVH is shown in Figure 5. Fine rows with a small corrugation can be seen extending along the main diagonal of the image. They correspond to the Si dimer rows. On top of the rows, the oligomers appear as elongated features. From this figure, they seem to be adsorbed randomly on the surface with no peculiar orientation in regard to the Si dimer rows. Even though many conformations are observed, the features have all almost the same lengths.

Using the dimer rows as a template, we find a length of 25 ± 3 Å, close to the 22 Å calculated distance between the outermost side carbon atoms of the conjugated chain. As the hexyl chains are much shorter (~9 Å), the features observed in Fig. 5 correspond to the backbones of the 4TVH oligomers. This result is in agreement with previous STM investigations of alkylated oligothiophenes adsorbed on graphite, where only the oligothiophene backbone was imaged [9].

Due to their saturation, alkanes form an insulator layer, when they adsorb on the silicon surface. From photoconductivity experiments of self-assembled monolayers of long alkanes chains, deposited on the silicon surface, it was found that the highest occupied molecular orbital (HOMO) of alkanes containing between 12 and 18 carbon atoms was 4 eV below the top of the silicon valence band [10]. At a negative sample voltage $V_0$, electrons tunnel from the valence band states of the semiconductor and the states of the molecules, if these states are lying in the energy range $eV_0$. As the difference between the top of the valence band and the HOMO of the alkanes increases when the alkane length decreases [11], this state for an hexyl chain is lying much below $eV_0$, at a voltage of -2.0 volts. Therefore, it does not contribute to the tunneling current. The hexyl chains can not be imaged in the voltage range commonly used while tunneling on semiconductors.

Although the alkyl chains do not appear in the STM images, it was shown that they played a role in the arrangement of decithiophenes adsorbed on graphite [9]. In our case, it is clear from the comparison between Fig. 3 and Fig. 5, which were acquired with the same sample voltage, that 4TV and 4TVH do adsorb in a different manner. While the 4TV appear as small adsorbates on the surface, the 4TVH are lying on the surface. Since the difference of structure between both molecules comes from the substitution of alkyl chains to the thiophene rings, we thus conclude that the difference of appearance in the STM images is caused by the alkyl chains.

To better understand the influence of the alkyl chains on the molecular arrangement of 4TVH, we have acquired STM images with different negative sample biases. Figures 6(a) and (b) were acquired simultaneously with two different voltages. A brief comparison between Fig. 6(a) and (b) reveals that the observation of the 4TVH oligomers in the STM images is highly voltage dependent. At sufficient high negative voltages, the molecules are visible, whereas at lower voltages, most of the molecules disappear and, only for a few molecules, their brightest part still remains visible. The calculated affinity and ionization energies of 4TV oligomers are respectively -1.75 eV and -5.79 eV [5]. The optical gap of 4TVH was found to be 2.4 eV [4]. Depending of the degree of coupling between the molecules and the surface, the HOMO of the 4TVH is thus positioned between 0.6 eV and 1.4 eV below the top of the silicon valence band, resonant with the occupied states of this band, as shown in figure 7. At high negative voltages, the electrons can then tunnel from the occupied states of the semiconductor and the HOMO of the oligomers to the empty states of the tip, thus allowing the appearance of the oligomers. Alternatively, at low negative voltages, only the states close from the Fermi level of the semiconductor can contribute to the tunneling current. These states, positioned in the silicon bulk gap, correspond only to the surface states associated with the Si dimers of the surface and therefore an STM image acquired at low negative voltages shows mainly the Si dimers of the surface. To emphasize this phenomenom, we can focus on the small row of ad-dimers [12] seen in the center of Fig. 6(a) anf Fig. 6(b), and perpendicular to the Si dimers rows of the surface. These dimers are lying on the surface and are thus positioned on the same plane as the molecules. They are bright in Fig. 6(a) and, in spite of the reduced voltage in Fig. 6(b), keep their brightness, in clear contrast with what can be observed for the molecules.

The electronic interaction between a molecule and a conductive electrode leads to the extension of the electrode wave functions into the molecule. As a result, the molecular levels are broadened. This high degree of broadening allows the appearance of molecules, adsorbed on metals, at low voltages through a virtual resonance tunneling process [13]. As far as the Si(100) surface is concerned, surface states are lying in the gap of the material and these states would allow a coupling with a broadened HOMO state. As most of the molecules does not appear at low biases, their HOMO state is therefore not coupled with the surface states in the band gap. We believe that the primary reason for the disappearance of the molecules, at low voltages, is caused by the alkyl chains. The dissociation of hexyl chains, which are saturated compounds like ethane, is not expected. As the hexyl do not bond to the Si(100) surface, steric hindrance prevents the oligomer backbone from reacting with the surface.

The weak interaction between 4TVH oligomers and the Si surface is supported by the observation of fuzzy oligomers, pointed by an arrow in Fig. 6(a). Such fuzzyness can be attributed to the displacement of the oligomers with the STM tip, which indicates a small interaction between the oligomers and the silicon surface.

While most of the oligomers seem to be physisorbed, a few oligomers show a very bright part at both voltages. Their interaction with the surface may be stronger. In this case, the hexyl chains would not isolate the entire backbone from the surface, allowing the unsaturated part of the oligomer to react with the Si dangling bonds. As a result, the observation of different adsorption configurations suggests that the arrangement of the hexyl chains is random but sufficient to prevent the 4TVH oligomers from forming a well ordered layer.

## IV. CONCLUSION

The molecular arrangement of 4TV and 4TVH oligomers adsorbed on the Si(100) surface has been studied by scanning tunneling microscopy. While the adsorption of 4TV oligomers on the Si(100) surface leads to the dissociation of the molecules, 4TVH keeps their structure intact after adsorption. This different behavior can be attributed to the steric effect of the hexyl chain which prevent the direct interaction of the π-conjugated system with the Si atoms. Thus, substitution of the thiophene rings by alkyl chains provides a possible way to position the oligomer with its main axis parallel to the surface, while isolating the conjugated backbone. In the frame of using such oligomers as molecular wires, the adsorption of 4TVH is interesting since the hexyl chains allow the molecules to preserve extended π-electronic delocalization. Furthermore, due to the low intrinsic solubility of the rigid conjugated chains, the synthesis of longer nTV oligomers, up to the decamer stage, required their substitution with alkyl chains. Therefore, the substitution with alkyl chains of longer nTV oligomers is very beneficent for their synthesis as well as for their use as potential molecular wires on the silicon surfaces. However the configuration adopted by the chains on the surface can vary from one oligomer to another and may make it difficult to selectively connect one end of the molecule to the surface. There is thus a need to substitute this end with functional groups capable of selectively react with the surface silicon atoms.

FIGURE CAPTIONS

Figure 1: Chemical structure of 4TV oligomers (a) and 4TV-H oligomers bearing hexyl chains at the β positions of the thiophene rings (b).

Figure 2: Electronic absorption spectra of the 4TV in a $CH_2Cl_2$ solution measured before and after the evaporation.

Figure 3: STM image of the Si(100) surface after deposition of 4TV oligomers. The image was acquired with a sample bias of -2.0 V and a tunneling current of 60 pA. Two different types of protusions, labelled A and B, can be seen. The grey scale ranges from 0 (black) to 3.2 Å (white).

Figure 4: High resolution STM image of the Si(100) surface after deposition of 4TV oligomers. The image was acquired with a sample bias of −2.0 V and a tunneling current of 40 pA.

Figure 5: STM image of the Si(100) surface after deposition of 4TVH oligomers. The image was acquired with a sample bias of −2.7 V and a tunneling current of 60 pA. The grey scale ranges from to 0 (black) to 2.6 Å (white).

Figure 6: Voltage dependent STM images of the Si(100) surface after deposition of 4TVH oligomers. The sample bias was in (a) -2.1 V and in (b) -1.3 V. The arrows indicate some fuzzy molecules. Three Si dimers positioned on top of the surface can be seen in the center of both images.

Figure 7: Schematic view of the energy diagram for semiconductor-molecule-vacuum-metal tunneling. CB and VB correspond to the conduction and valence band of the semiconductor. The semiconductor Fermi level is denoted $E_F$.

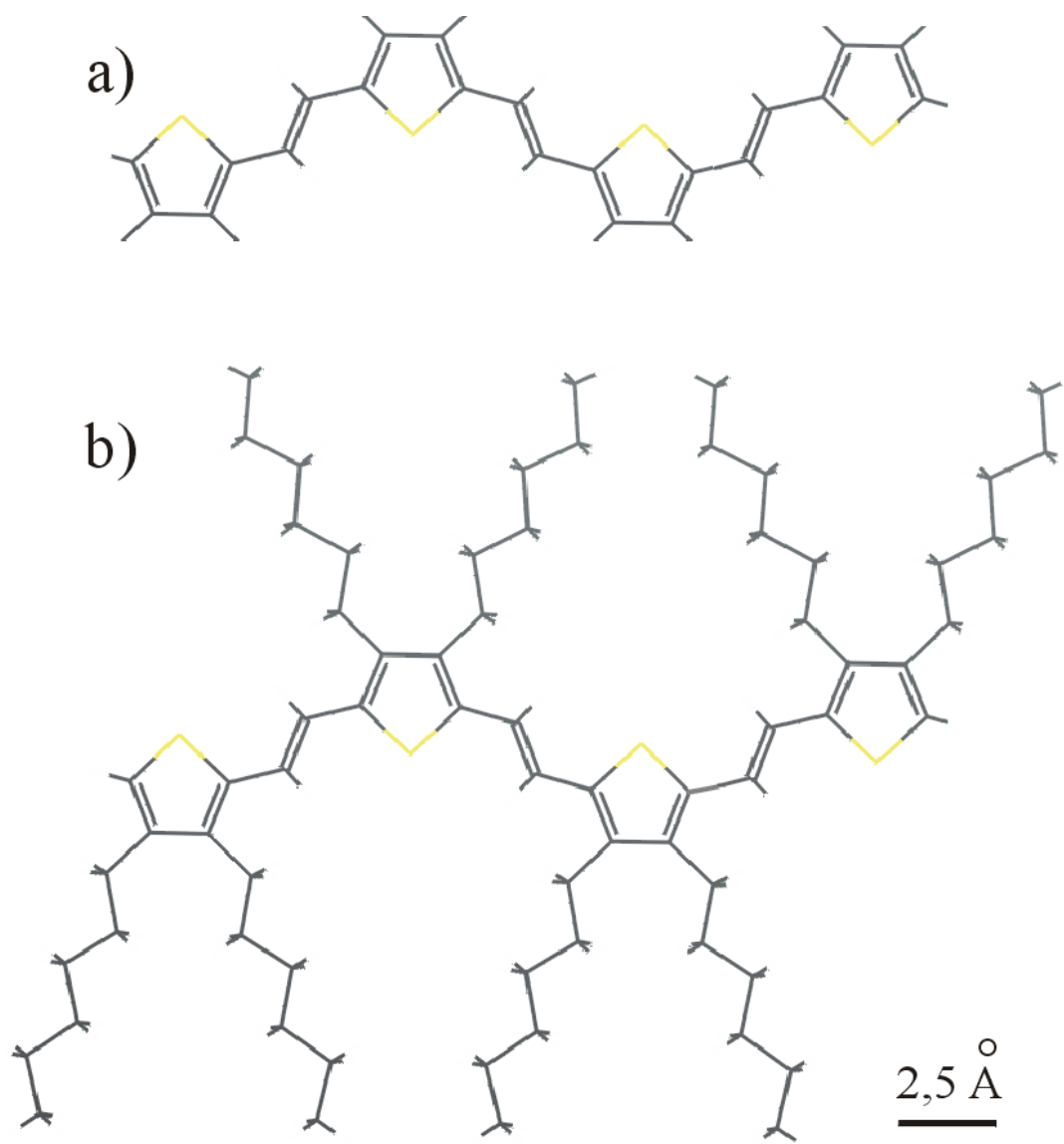

Figure 1

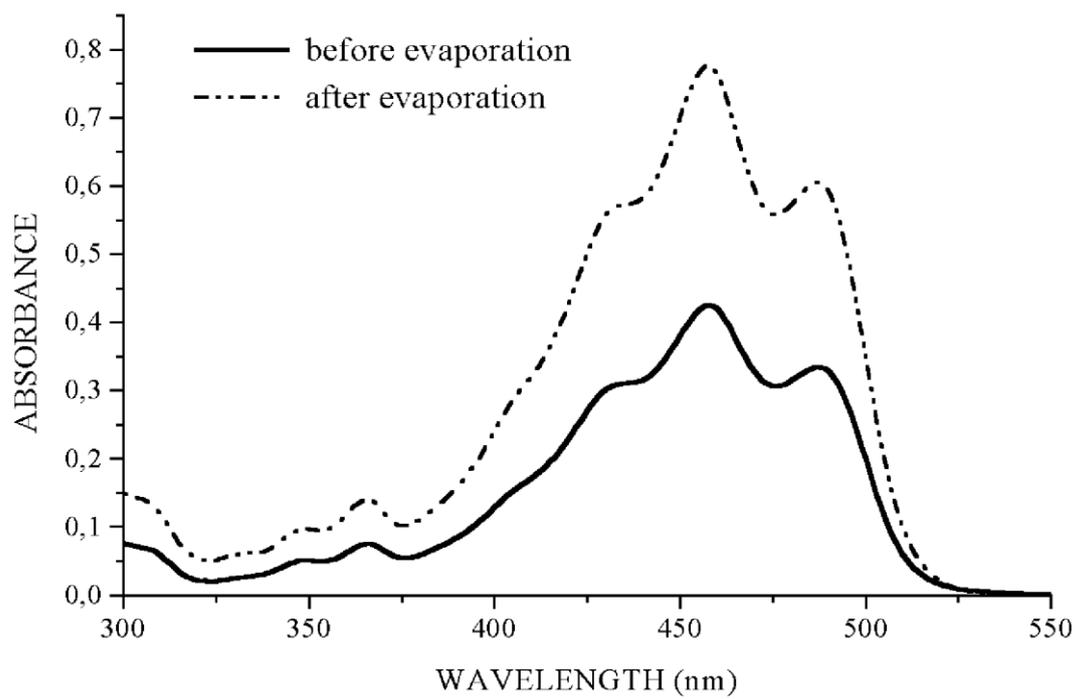

FIGURE 2

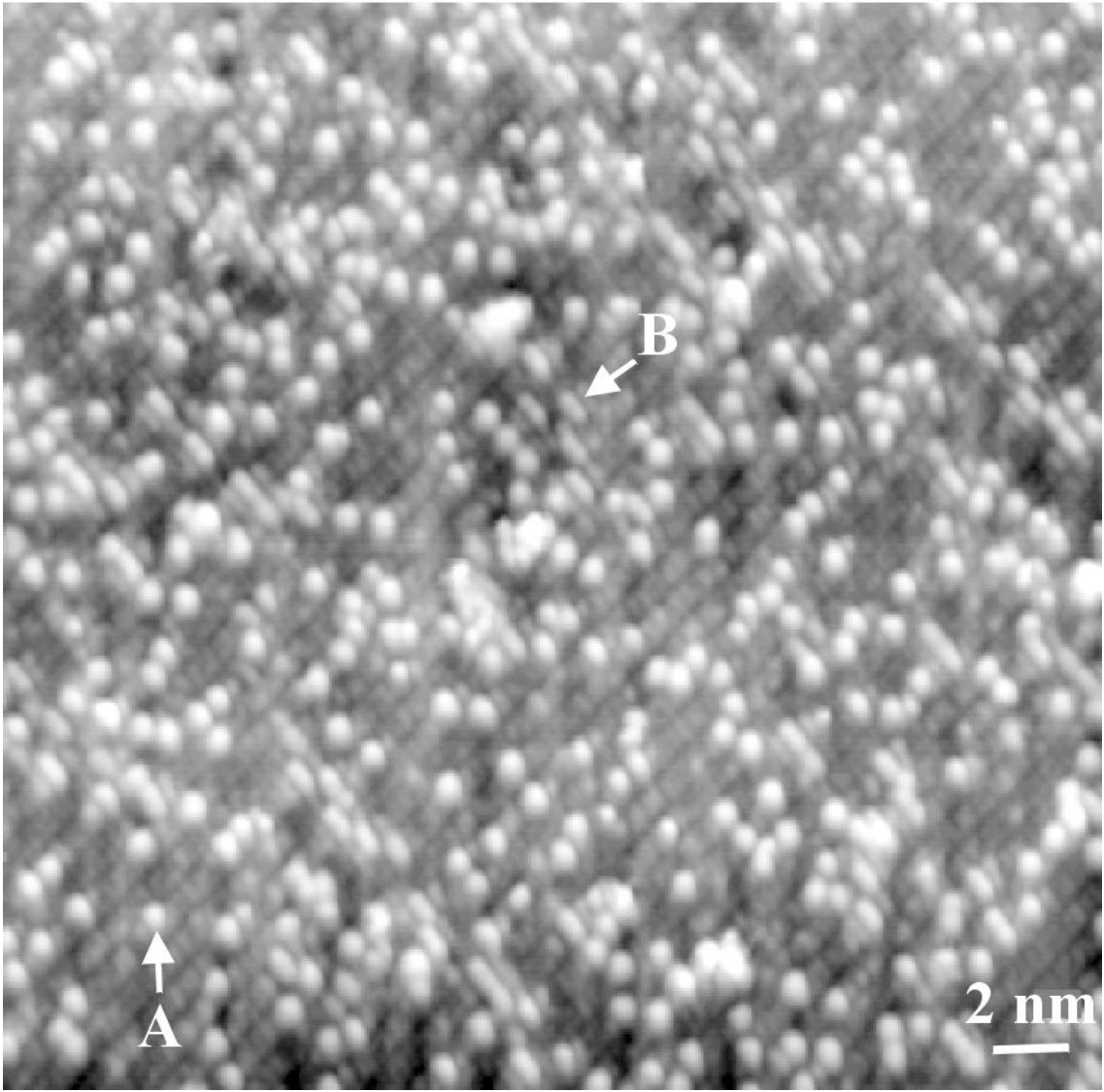

Figure 3

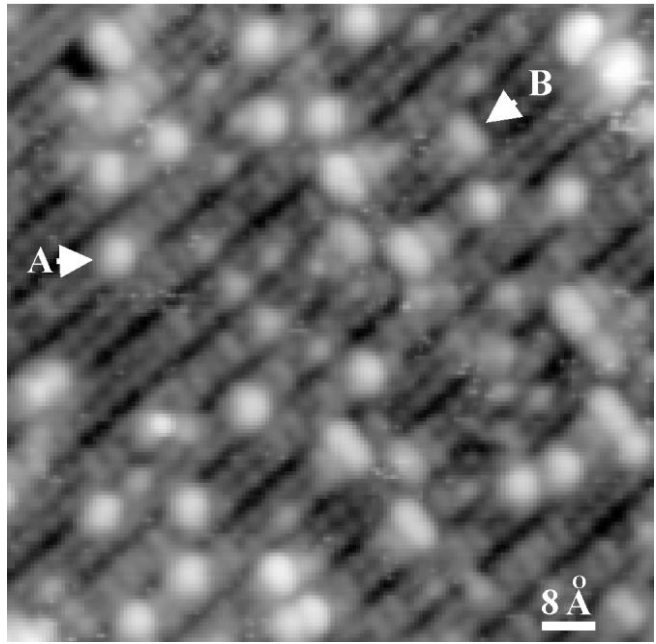

Figure 4

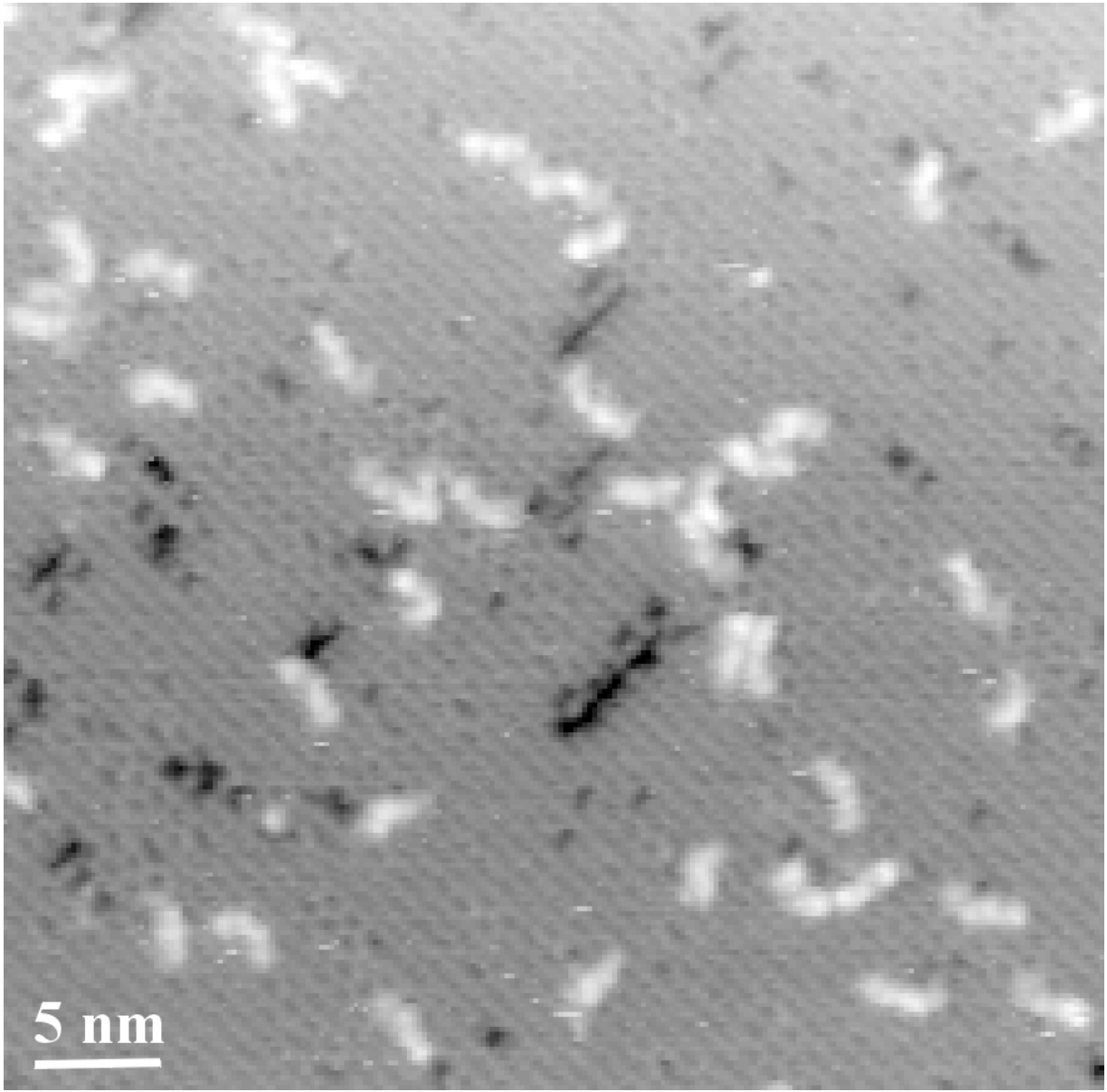

Figure 5

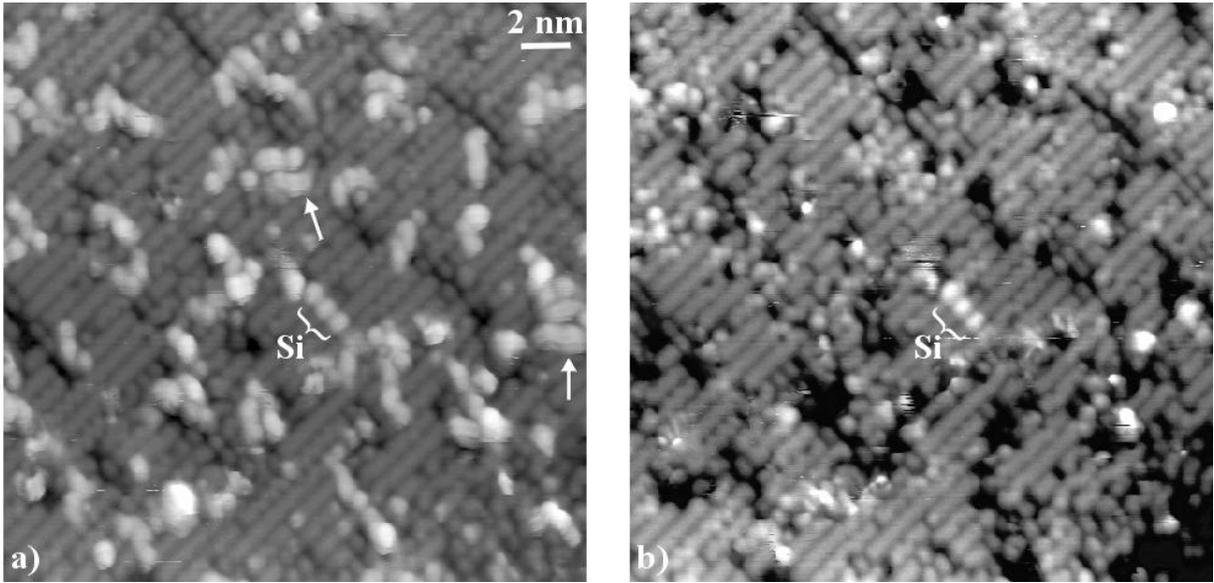

Figure 6

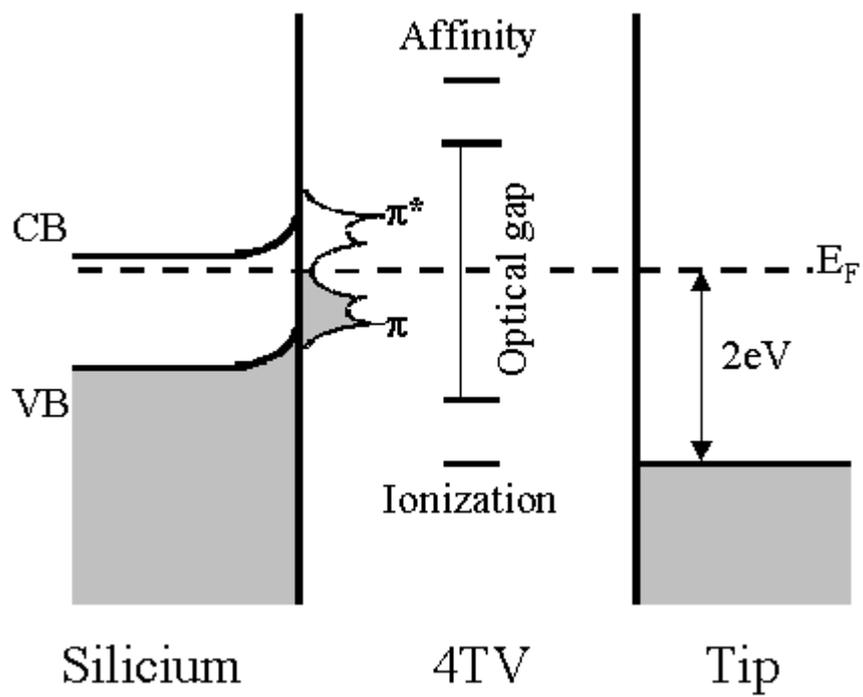

Figure 7